# Ultra-Long Homochiral Graphene Nanoribbons Grown Within *h*-BN Stacks for High-Performance Electronics

Bosai Lyu[1,2,†], Jiajun Chen[1,2,†], Sen Wang[3,4,†], Shuo Lou[1,2,†], Peiyue Shen[1,2,†], Jingxu Xie[1,2,†], Lu Qiu[5,6,7,†], Izaac Mitchell[5,†], Can Li[1,2], Cheng Hu[1,2], Xianliang Zhou[1,2], Kenji Watanabe[8], Takashi Taniguchi[9], Xiaoqun Wang[1,2,10], Jinfeng Jia[1,2,10], Qi Liang[1,2,10], Guorui Chen[1], Tingxin Li[1,10], Shiyong Wang[1,2,10], Wengen Ouyang[3,4,*], Oded Hod[11], Feng Ding[5,12,*], Michael Urbakh[11,*], Zhiwen Shi[1,2,10]*

[1]Key Laboratory of Artificial Structures and Quantum Control (Ministry of Education), Shenyang National Laboratory for Materials Science, School of Physics and Astronomy, Shanghai Jiao Tong University, Shanghai 200240, China.

[2]Collaborative Innovation Center of Advanced Microstructures, Nanjing University, Nanjing 210093, China.

[3]Department of Engineering Mechanics, School of Civil Engineering, Wuhan University, Wuhan 430072, China.

[4]State Key Laboratory of Water Resources and Hydropower Engineering Science, Wuhan University, Wuhan 430072, China

[5]Department of Materials Science and Engineering, Ulsan National Institute of Science and Technology, Ulsan 44919, South Korea

[6]Graduate School of Carbon Neutrality, Ulsan National Institute of Science and Technology, Ulsan 44919, South Korea

[7]Department of Mechanical Engineering, Ulsan National Institute of Science and Technology, Ulsan 44919, South Korea

[8]Research Center for Functional Materials, National Institute for Materials Science, 1-1 Namiki, Tsukuba 305-0044, Japan.

[9]International Center for Materials Nanoarchitectonics, National Institute for Materials Science, 1-1 Namiki, Tsukuba 305-0044, Japan.

[10]Tsung-Dao Lee Institute, Shanghai Jiao Tong University, Shanghai, 200240, China.

[11]Department of Physical Chemistry, School of Chemistry and The Sackler Center for Computational Molecular and Materials Science, The Raymond and Beverly Sackler Faculty of Exact Sciences, Tel Aviv University, Tel Aviv 6997801, Israel.

[12]Shenzhen Institute of Advanced Technology, Chinese Academy of Sciences, Shenzhen 518055, China

[†]These authors contributed equally to this work.

*Correspondence to: w.g.ouyang@whu.edu.cn, f.ding@unist.ac.kr, urbakh@tauex.tau.ac.il, zwshi@sjtu.edu.cn

**Van der Waals encapsulation of two-dimensional materials within hexagonal boron nitride (*h*-BN) stacks has proven to be a promising way to create ultrahigh-performance electronic devices.[1-4] However, contemporary approaches for achieving van der Waals encapsulation, which involve artificial layer stacking using mechanical transfer techniques, are difficult to control, prone to contamination, and unscalable. Here, we report on the transfer-free *direct growth* of high-quality graphene nanoribbons (GNRs) within *h*-BN stacks. The as-grown embedded GNRs exhibit highly desirable features being ultralong (up to 0.25 mm), ultranarrow (< 5 nm), and homochiral with zigzag edges. Our atomistic simulations reveal that the mechanism underlying the embedded growth involves ultralow GNR friction when sliding between AA′-stacked *h*-BN layers. Using the grown structures, we demonstrate the transfer-free fabrication of embedded GNR field-effect devices that exhibit excellent performance at room temperature with mobilities of up to ~4,600 cm$^2$V$^{-1}$s$^{-1}$ and on-off ratios of up to ~10$^6$. This paves the way to the bottom-up fabrication of high-performance electronic devices based on embedded layered materials.**

Graphene, a two-dimensional (2D) crystal consisting of a single-layer of carbon atoms arranged in a honeycomb lattice, has been intensively investigated since its first isolation in 2004.[5-9] As a candidate for future electronic materials, although with ultrahigh carrier mobility, pristine graphene suffers from the absence of an electronic bandgap. One may overcome this problem by considering graphene nanoribbons (GNRs) - quasi-one-dimensional graphene stripes possessing finite bandgaps due to quantum confinement.[10,11] Theoretically, sub-5 nm-wide pristine GNRs are predicted to exhibit large bandgaps that are suitable for room-temperature on-off operation, accompanied by charge carrier mobilities as high as 10,000 $cm^2V^{-1}s^{-1}$.[12,13] This makes them ideal platforms for digital logic and radio-frequency (RF) electronic applications. In practice, however, significantly lower mobilities (< 1500 $cm^2V^{-1}s^{-1}$) have been observed for prototypical GNR-based field-effect transistors (FETs).[14-24] This discrepancy between theoretical predictions and experimental findings can be attributed to disorder effects,[25] including lattice defects, strain, surface roughness, physical and chemical adsorption of contaminants, and substrate-charged impurities. These disorder effects are of special significance due to the low-dimensional nature of GNRs, possessing only surfaces and edges.

To enhance the performance of graphene-based devices, various methods have been introduced for reducing disorder effects, including thermal annealing,[26] plasma[27] and atomic force microscopy (AFM)[28] surface cleaning, fabrication of suspended architectures,[29] polymer-based transfer to flat substrates, and van der Waals (vdW) encapsulation.[30] The most successful approach so far is vdW encapsulation of graphene between stacks of hexagonal boron nitride ($h$-BN) – an atomically flat wide-bandgap-layered insulator. Encapsulated 2D graphene devices exhibit ultralow charge inhomogeneity ($\delta n$ ~$10^9$ $cm^{-2}$), ultra-high carrier mobilities (~$10^6$ $cm^2V^{-1}s^{-1}$), and ultralong mean free paths (~100 μm). Nevertheless, the encapsulation procedure involves mechanical vdW assembly – a low-yield technique that produces only micrometer-scale samples, making it unsuitable for advanced electronic applications.[1]

In the present study, we introduce a new scalable synthetic approach for the direct growth of GNRs, embedded within insulating $h$-BN stacks. The as-grown embedded-GNRs are as long as ~250 μm, ultranarrow (<5 nm), and feature uniform zigzag edge geometry. The growth mechanism involves ultralow friction GNR sliding between AA′-stacked $h$-BN layers, resulting from effective cancellation of the interactions between the ribbon atoms and the surrounding insulator layers. This, in turn, defines a preferred nearly free sliding path dictating the chirality of the grown GNRs. FETs fabricated using our samples exhibit superior electronic characteristics including room-temperature carrier mobilities of up to ~4,600 cm$^2$V$^{-1}$s$^{-1}$, on-off ratios of up to ~10$^6$, and subthreshold swings of ~100 mV dec$^{-1}$. This reflects the effectiveness of the *in-situ* encapsulation process and the high quality and cleanness of the grown embedded GNRs.

## Growth and structural characterizations

The growth process of embedded GNRs is briefly illustrated in Fig. 1a. Experimentally, iron nanoparticles are first deposited onto multi-layered *h*-BN flakes residing on top of an SiO$_2$/Si substrate. The system is then heated in a chemical vapour deposition (CVD) tube furnace under an atmospheric pressure flow of H$_2$ and Ar mixture (see *Methods*). The heating induces nanoparticle migration towards the *h*-BN step edges (Extended Data Fig. 1), such that the catalytic surface gets exposed to the void interlayer regions within the stack. At a growth temperature of ~850 °C, methane is introduced as the carbon source, initiating embedded GNR growth at the nanoparticle surface into one of the interlayer gaps (see Fig. 1a, b and Extended Data Fig. 2). The leading edge of the growing GNR is pushed forward into the interlayer region as additional carbon rows are added to its trailing edge at the nanoparticle surface. Following the growth stage, the sample is cooled down to room temperature under the protective H$_2$/Ar atmosphere. The catalytic growth process is similar to that recently proposed for GNR growth atop *h*-BN surfaces.[31] Since some GNRs also grow atop the upper *h*-BN surface, the sample is then treated with hydrogen-plasma etching, leaving only the embedded GNRs (Extended Data Fig. 3). Further details regarding the growth process can be found in the *Methods* section.

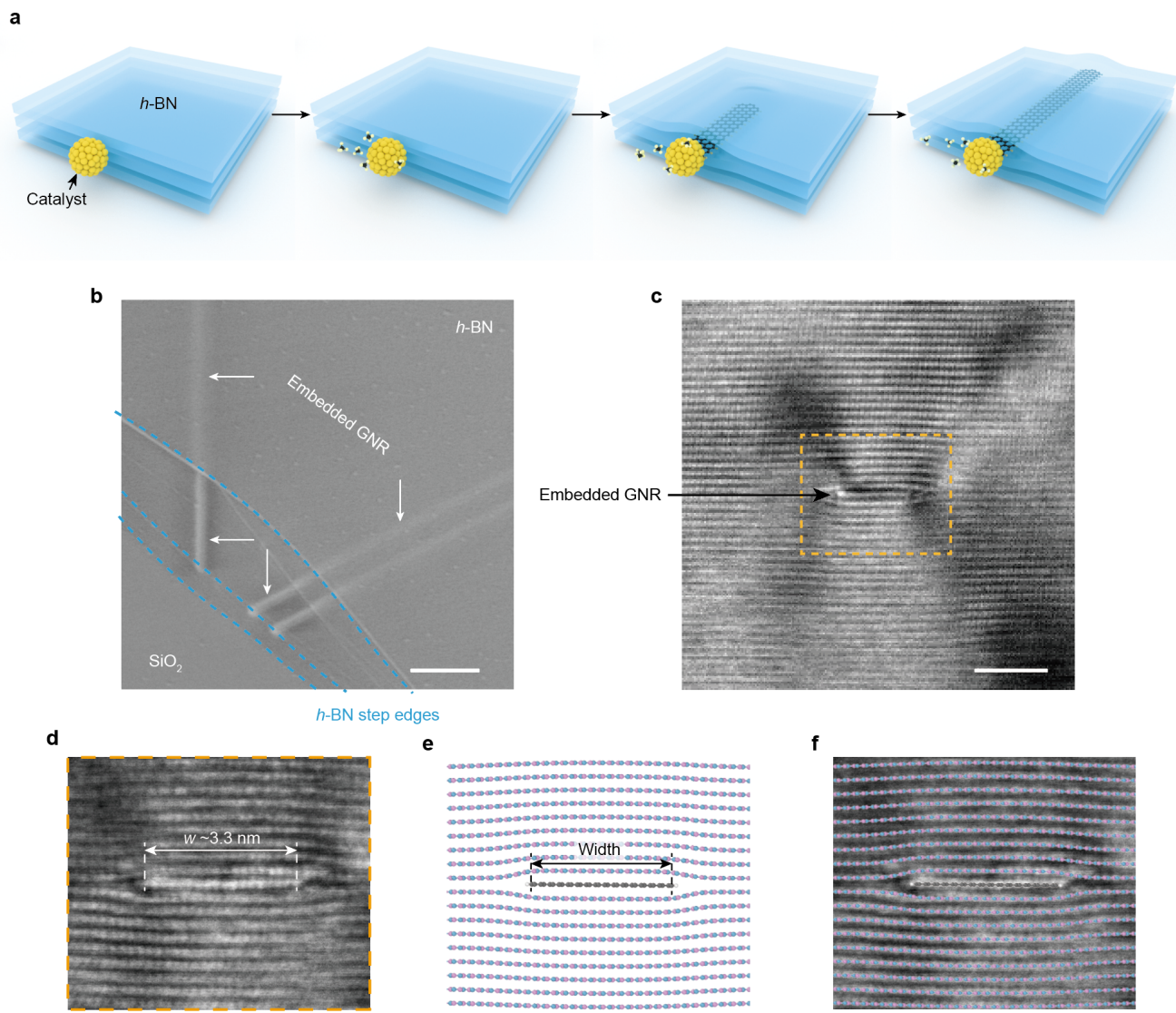

**Figure 1. Directly-grown embedded graphene nanoribbons. a,** Schematic of the growth process of an *h*-BN embedded GNR. The growth is catalysed by an Fe nanoparticle attached to an *h*-BN step edge. **b,** SEM image of the as-grown GNRs (bright straight lines) embedded within an *h*-BN stack. The GNRs contrast blurs with increasing *h*-BN cover thickness (see dashed blue lines representing step edges separating regions of different *h*-BN stack thickness). Scale bar, 1 μm. **c,** Cross-sectional STEM (200 kV) dark-field image of an as-grown embedded GNR. Scale bar, 3 nm. **d,** Magnification of the region marked by the orange dashed rectangle in panel (c). Each bright spot represents an axial view of a zigzag chain of the hexagonal lattice of a GNR or an *h*-BN layer. The GNR shows a somewhat higher contrast than the insulating *h*-BN. **e,** Atomistic representation of the cross-section of a 3.3 nm-wide GNR embedded within a 40-layer *h*-BN stack, as obtained using classical force-field calculations. **f,** Overlaid cross-sectional image combining the experimental image of panel (d) and the calculated structure of panel (e).

Cross-sectional views of encapsulated GNRs are obtained using scanning transmission electron microscopy (STEM, see *Methods*). In the STEM image in Fig. 1c, the cross-section of a 3.3-nm wide (see magnified view in Fig. 1d) monolayer GNR is clearly seen perturbing the otherwise pristine *h*-BN layered stack, where each bright spot (seen better in Fig. 1d) represents an axial view of a zigzag graphene or *h*-BN chain. Similar images of additional embedded GNRs, 3-5 nm wide, are shown in Extended Data Fig. 4. To accommodate the embedded GNR, the *h*-BN stack deforms vertically with an elastic penalty that is compensated by increasing attractive interlayer dispersive interactions. The *h*-BN layers adjacent to the GNR, demonstrate localized deformation near the GNR edges, whereas at the surface region they flatten to maximize the vdW interactions with the planar GNR. Notably, the vertical deformation propagates into the *h*-BN bulk with a relatively high penetration length.[32] However, away from the GNR, the deformation becomes delocalized and the *h*-BN layers develop an arc-like structure. These experimental observations are fully captured by our atomistic classical force-field calculations (see *Methods*) presented in Figs. 1e and Fig. S1 of the Supplementary Information (SI). The remarkable agreement is clearly demonstrated by overlaying the experimental and computational results as shown in Fig. 1f.

Scanning electron microscopy (SEM, see *Methods*) is used to provide a top-view image of the fully grown samples (see Fig. 2a, b and Extended Data Fig. 5). The embedded GNRs demonstrate high contrast against the insulating *h*-BN host, with increasing blurriness as the thickness of the *h*-BN cover increases (see, e.g., Fig. 1b). The embedded GNRs are found to be perfectly straight with a length ranging from several to a few hundred micrometres (Fig. 2a-c), much longer than on-surface grown GNRs (see inset of Fig. 2c). Notably, the aspect ratio of the longest embedded GNRs obtained (~250 μm) is ~$1\times10^5$, at least two orders of magnitude larger than GNRs synthesized by other methods so far[16,33-37] (see Fig. 2d and SI Table S1). Furthermore, the embedded GNRs grow mainly along the zigzag axes of the *h*-BN stack (see Figs. 2b and Extended Data Fig. 5), where longer GNRs show higher tendency to align. Notably, above a length of 20 μm they are all found to grow along one of the zigzag axes of the *h*-BN stack (Fig. 2e).

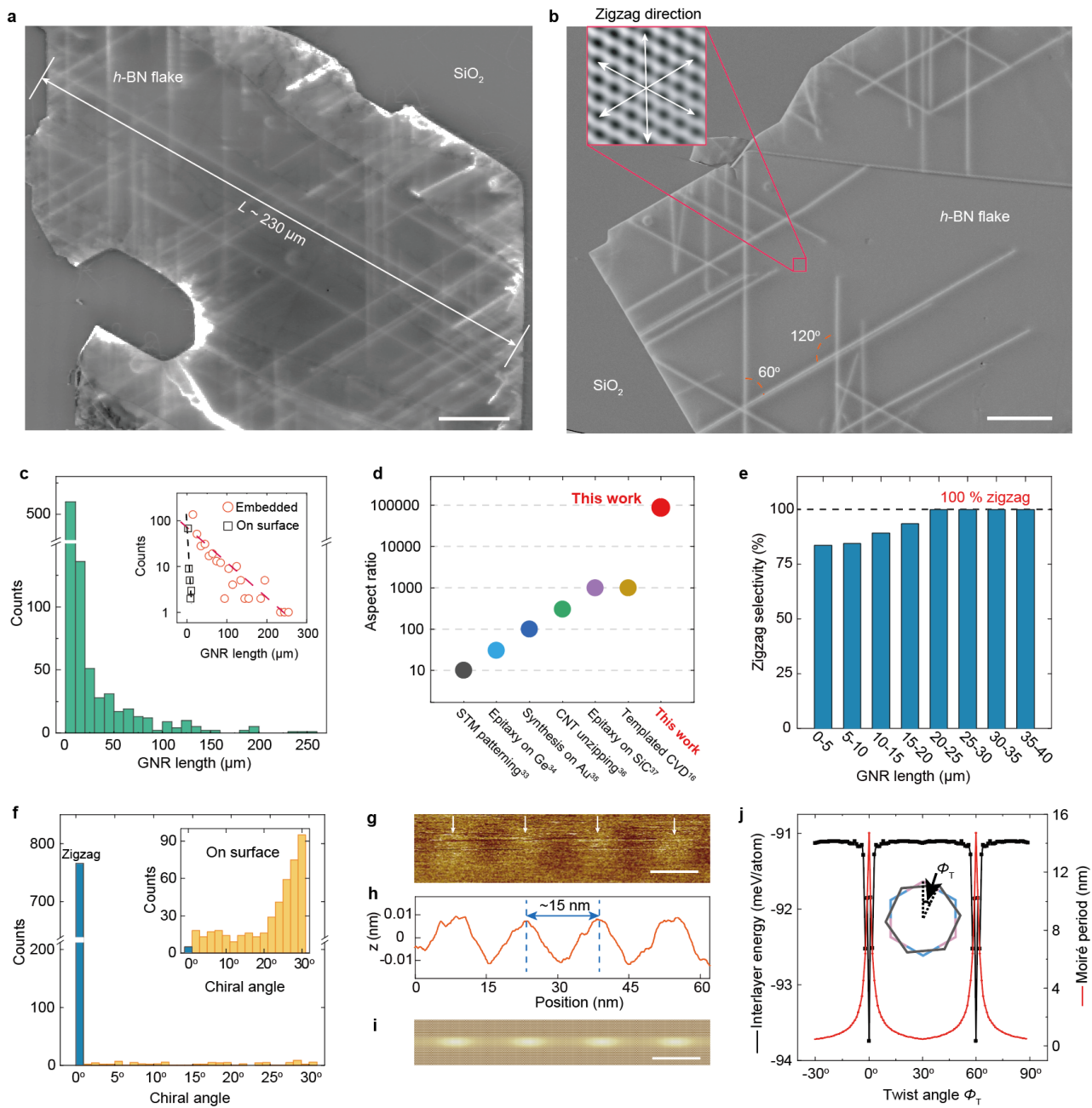

**Figure 2. Ultralong embedded zigzag GNRs exhibiting 1D moiré superstructures. a**, A large-scale SEM image of embedded GNRs with lengths up to a few hundreds of micrometres. Scale bar, 30 µm. **b**, SEM image demonstrating that embedded GNRs grow mainly along the zigzag *h*-BN axes. Inset, *h*-BN surface crystal directions measured using friction-mode AFM. The white arrows denote the zigzag axes. Scale bar, 5 µm. **c**, Length histogram of embedded GNRs. Inset, a semi-log plot comparing the length distribution of embedded and on-surface-grown GNRs demonstrating that the former are substantially longer. **d**, Comparison of the aspect ratio of GNRs synthesized using different approaches. **e**, Correlation between embedded GNR length and the selectivity of their growth direction along the *h*-BN stack. **f**, Chirality distribution of the *h*-BN embedded GNRs, showing a strong preference towards the growth of ZZ-GNRs. Inset, chirality distribution of on-surface grown GNRs showing preference to the growth of AC-GNRs but with a significantly wider distribution. **g**, High-magnification AFM topography image of the *h*-BN surface atop an embedded zigzag GNR, showing clear 1D periodic moiré patterns along the GNR. Scale bar, 10 nm. **h**, Height profile extracted from the 1D moiré superlattice in panel (g), showing a periodicity of ~15 nm and a corrugation of 20 pm. **i**, Calculated topography image of the one-dimensional (1D) moiré pattern along an embedded zigzag GNR, crystallographically aligned with the *h*-BN stack (see *Methods*). Scale bar, 10 nm. **j,** Calculated dependence of the embedded GNR/*h*-BN interlayer stacking energy (black curve) and the moiré period (red curve) on the twist angle (see SI section 4). At zero twist angle the stacking energy is minimal and the moiré period is maximal (~15 nm).

Owing to the long-range deformations induced by the embedded GNRs, *h*-BN surface topography AFM imaging is able to detect them, showing a maximal height corrugation of ~0.2 nm for near-surface GNRs, which reduces with embedded GNR depth (see SI Fig. S2 and Extended Data Fig. 1). Importantly, the *h*-BN surface AFM scans atop the embedded ribbons exhibit periodic features in the height profile (see Fig. 2g, h) with an amplitude of ~0.01 nm. The observed periodicity of ~15 nm (see Fig. 2h) matches that of the 2D moiré patterns emerging in aligned graphene/*h*-BN interfaces[38] due to their intrinsic interlayer lattice mismatch of ~1.8% (see Fig. 2i). This indicates that not only is the main axis of the embedded GNRs aligned with the zigzag directions of the *h*-BN stack, but they are also crystallographically matched with the hexagonal *h*-BN lattice. Namely, the grown GNRs are predominantly of zigzag type (see Fig. 2f), and hence manifest a one-dimensional (1D) moiré superstructure with the adjacent *h*-BN layers.[31] This can be rationalized considering the interlayer stacking energy, which obtains a minimum at zero twist angle between zigzag GNRs (ZZ-GNRs) and the embedding *h*-BN surfaces (see Fig. 2j). We note that the high zigzag edge selectivity obtained for embedded GNRs is in stark contrast to the case of their on-surface-grown counterparts (and most other GNR fabrication approaches), where a wide distribution of GNR chiralities is often obtained (see inset of Fig. 2f).

## Growth mechanism

As indicated by the experimental observations, the embedded GNR growth mechanism involves the following ingredients: (i) nucleation at the catalytic nanoparticle; (ii) addition of GNR segments at the surface of the nanoparticle, driving the leading GNR edge to slide away from the nucleation site; (iii) penetration of the GNR into the *h*-BN interlayer spacing; (iv) growth of the GNR within the *h*-BN stack. The growth process is governed by interlayer lattice commensuration effects, deformation energy penalty, increased vdW interactions, and frictional energy dissipation. The balance between these factors dictates the preferred growth orientation, edge type, and overall length of the embedded GNRs. Specifically, the growth is expected to terminate once the sliding friction force exceeds the growth driving force produced by the catalytic reaction at the surface of the nanoparticle. Therefore, superlubric sliding is expected to promote the growth of ultralong embedded GNRs. In this context, the fact that *h*-BN embedded GNRs grow typically longer than their on-surface counterparts is counterintuitive and deserves further exploration.

To this end, we performed fully atomistic molecular dynamics simulations (MD, see *Methods*) of the sliding motion of ZZ-GNRs and armchair GNRs (AC-GNRs) between and on top of *h*-BN layers (Fig. 3a). Figure 3b shows the penetration length of GNRs into an *h*-BN stack as a function of constant pushing force (mimicking the force exerted by the catalytic growth centre). Notably, in correspondence with the experimental observations, under a given pushing force embedded ZZ-GNRs (orange squares) propagate much further into the *h*-BN stack than embedded AC-GNRs (green circles) and on-surface ZZ-GNRs (blue triangles). To rationalize this intriguing finding, we consider first the energy landscapes of ZZ-GNRs sliding on top of (Fig. 3c) and within (Fig. 3d) an *h*-BN stack (see *Methods* section). Atop the *h*-BN surface, the low energy regions (blue, Fig. 3c) in the ZZ-GNR sliding potential energy surface are localized and separated by relatively high energy barriers (red, Fig. 3c). Upon sliding, the system is, therefore, forced to cross the energy barriers (see dashed blue line in Fig. 3c and blue triangles in Fig. 3f), leading to increased friction. When sliding between *h*-BN layers, however, a continuous low-energy valley develops (blue, Fig. 3d and a more detailed analysis is shown in SI Fig. S3) allowing for nearly frictionless sliding (orange line in Figs. 3d and orange squares in Fig. 3f) of the ZZ-GNR with minor sideways motion to avoid the localized high-energy peaks (red, Fig. 3d). This explains the fact that embedded ZZ-GNRs have a tendency to grow longer than their on-surface counterparts. Furthermore, examining the sliding energy profile of embedded AC-GNRs, a continuous low energy valley is also found to be present (blue, Fig. 3e). However, to avoid the high-energy peaks (red, Fig. 3e), considerably larger side-ways motion and ribbon deformation along the sliding trajectory are required (dashed green line in Fig. 3e and green circles in Fig. 3f), leading to the observed strong preference towards ZZ-GNR growth (Fig. 2f).

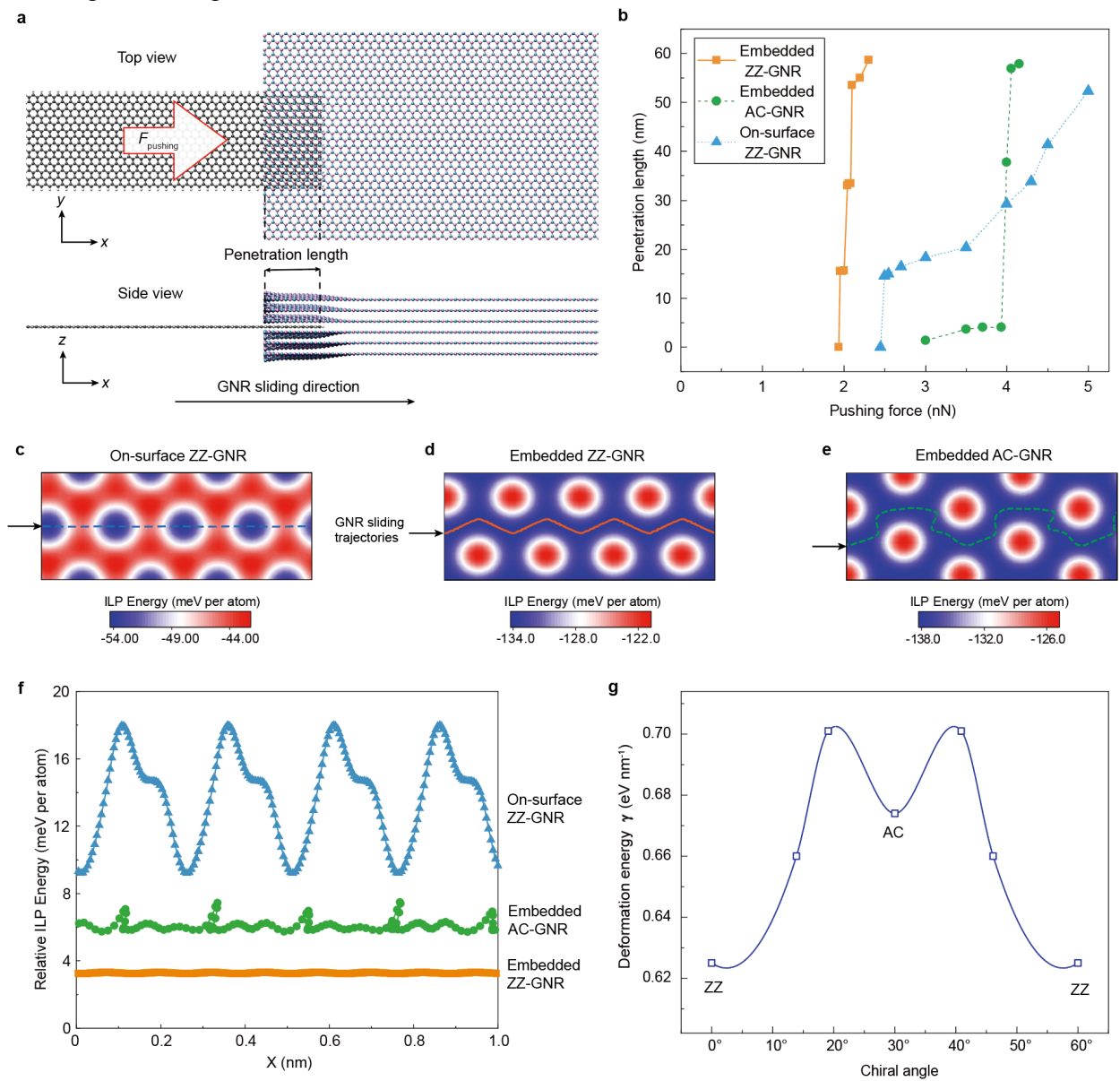

**Figure 3. Embedded GNR sliding mechanism. a,** Schematic of the simulation setup for ZZ-GNR penetration into an *h*-BN stack. **b**, Calculated penetration distances of ZZ-GNRs (orange squares) and AC-GNRs (green circles) inside an *h*-BN stack, as a function of applied pushing force. The corresponding results for on-surface ZZ-GNR sliding are presented by the blue triangles. Details regarding the calculations can be found in the *Methods* section. **c-e**, Sliding potential energy landscapes and trajectories for short ZZ-GNRs on (c) and within (d) the *h*-BN stack, and (e) for embedded AC-GNRs. **f**, Potential energy variations along the sliding trajectories marked in panels (c-e). **g,** *h*-BN deformation energy as a function of embedded GNR chirality. Deformation energy calculations details are given in SI section 6. AC, armchair, ZZ, zigzag.

Although the sliding energy barrier considerations favor the growth of embedded ZZ-GNRs over embedded AC-GNRs, we should also consider the effects of unavoidable vertical stack deformation during embedded growth. Figure 3g shows the deformation energy penalty for various GNR chiralities (see SI section 6 for further details), demonstrating that ZZ-GNRs induce lower deformation energy costs, thus further supporting the strong preference towards their growth. We, therefore, conclude that the overall effect of interlayer lattice commensuration, deformation energy penalty, vdW interactions, and frictional energy dissipation results in highly selective growth of ultralong embedded ZZ-GNRs.

## High-performance GNR FETs

The *h*-BN embedded GNRs allow for the fabrication of high-quality protected-channel devices, while circumventing the need for mechanical transfer processes and eliminating effects of oxidation, environmental contamination, and photoresist damage. To demonstrate this, we fabricated FET devices, based on the as-grown embedded GNRs (see Fig. 4a). Electrical contacts were made by reactive ion etching of the *h*-BN/GNR/*h*-BN heterostructure to expose the edge terminals of the GNR, followed by metal lead deposition (see *Methods,* inset of Fig. 4a). The resulting contacts are similar to the 1D contacts used to measure transport in mechanically transferred encapsulated 2D graphene.[1] Figure 4b shows a 2D colour plot of room-temperature current ($I_{sd}$) as a function of source-drain voltage ($V_{sd}$) and gate voltage ($V_g$). A typical diamond shaped map is obtained, where a low-current region (blue) of $\Delta V_g \sim 15$ V is obtained at the low source-drain bias voltage corresponding to the case where the Fermi energy lies within the bandgap of the ribbon. This is in agreement with the semiconducting nature of ultrathin ZZ-GNRs.[10,11,31] The low-current region narrows with increasing bias voltage, manifesting the larger Fermi transport window and the effect of the bias on the band alignment diagram of the device. This behaviour is indicative of a Schottky barrier FET device,[39,40] as was observed in several ultrathin FETs, especially in nanomaterials that exhibit small bandgaps (see SI section 9 for further details).[39] A few line-cuts of the 2D plot at representative $V_{sd}$ and $V_g$ are shown in Figs. 4c and 4d demonstrating a high on-off ratio of $I_{on}/I_{off} \sim 10^6$. Similar behaviour is obtained for other embedded GNR FET devices, which exhibit on-off ratios in the range $10^3$ - $10^6$ (see Extended Data Fig. 6). Given that $I_{on}/I_{off} \propto \exp(E_g/2k_BT)$, where $k_B$ is the Boltzmann's constant and $T$ is temperature, the measured on-off current ratio range corresponds to a bandgap $E_g$ range of 0.2 - 0.6 eV, typical of the experimentally observed GNR width range of 2-5 nm, consistent with our previous measurements of on-surface-grown GNRs bandgaps.[14,31] Note that the semiconducting nature of GNRs, which differs from carbon nanotubes (CNTs) that may exhibit metallic character, is important for creating nanoelectronic devices and circuits. (See SI section 11). Among all devices fabricated, the maximum output current measured exceeded 8 µA (see SI Fig. S6).[41,42] Moreover, the field-effect carrier mobility $\mu = \frac{dG}{dV_g} \times L/C_{gs}$ of the embedded-GNR devices falls in the range 1,400-4,600 cm$^2$ V$^{-1}$s$^{-1}$ (see *Methods* and Extended Data Figs. 7a-e), where $G$ is the measured conductance of the FET device, $L$ is the channel length, and $C_{gs}$ is the effective capacitance. The variation in carrier mobilities obtained for different devices can be attributed to differences in the electrical contact quality at the terminals of the ultranarrow GNRs (see SI Fig. S7) and variability in the intrinsic bandgaps of the GNRs. Nonetheless, the highest room-temperature mobility that we measured constitutes a record value for narrow GNR-based devices (the red star in Fig. 4e). Furthermore, at a temperature of 10 K the mobility of this GNR device reaches an exceptionally high value of ~74,000 cm$^2$ V$^{-1}$s$^{-1}$ (see Extended Data Fig. 8). These findings reflect the low defect density and high homogeneity of the as-grown embedded GNR samples. These two merits give rise to a rather small subthreshold swing of ~100 mV dec$^{-1}$ (see *Methods* and Extended Data Figs. 7f and 7g), which is close to the theoretical limit of 60 mV dec$^{-1}$ at room temperature as determined by thermal excitation.[13,43,44] Notably, these excellent FET characteristics are obtained at room temperature, demonstrating the suitability of our as-grown embedded GNRs to serve as active components in nanoscale electronic devices. Finally, we note that, high-performance GNR device with low-resistance Ohmic contacts and a shorter channel length comparable to the mean free path would also be beneficial to explore ballistic or quasi-ballistic transport phenomena, such as quantum conductance and Fabry-Perot interference.[45-48]

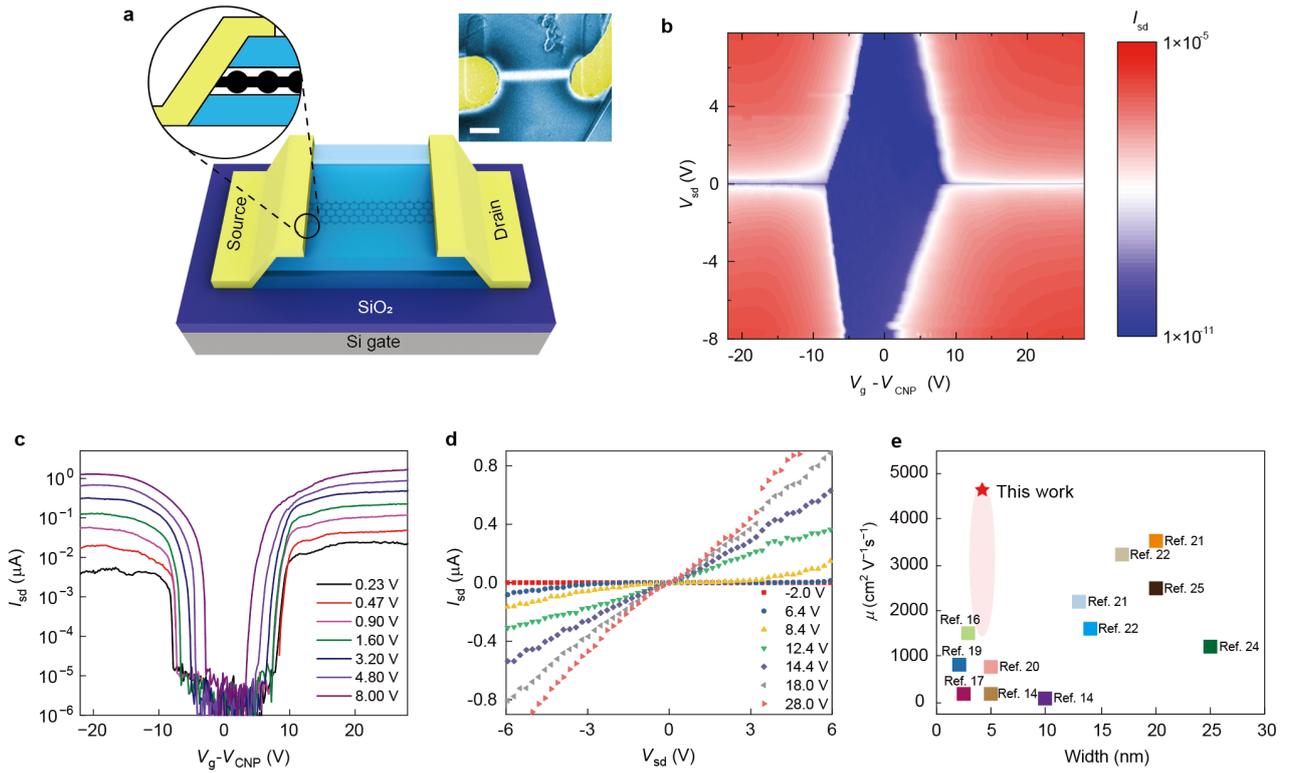

**Figure 4. Superior field-effect transistors based on the embedded-GNRs. a**, Schematic of a FET device made of an embedded-GNR. Inset, false-coloured SEM image of a typical embedded-GNR FET device with two Au/Cr source-drain electrodes. Scale bar, 2 μm. **b**, 2D colour plot of the source-drain current (log $I_{sd}$) as a function of source-drain voltage $V_{sd}$ and gate voltage $V_g$-$V_{CNP}$ at room temperature for a single-layer GNR device of channel length of ~17 μm and width of ~3 nm, where gate voltage for the charge neutrality point ($V_{CNP}$) is -28V (see SI section 8). **c**, Line-cuts at different source-drain voltages along the gate voltage axis in panel (b), yielding the device transfer characteristics, with notable on-off ratios of ~$10^6$. **d**, The output characteristics extracted from panel (b). **e**, Comparison of the embedded GNR room-temperature carrier mobility (pink region) to state-of-the-art results found in the existing literature. The red star marks the best result obtained for embedded GNR devices fabricated in this study.

## Discussion

Going beyond devices based on individual GNRs, the extension of our catalytic growth approach towards multi-ribbon architectures may pave the way to the design and fabrication of novel GNR-based nanoelectronic components. As an example, Extended Data Fig. 9 presents AFM images of three typical multi-ribbon configurations obtained in our experiments: (i) parallel GNR pairs grown at different *h*-BN stack depth (see Extended Data Fig. 9a) allowing the study of Coulomb drag effects[49]; (ii) X-shaped crossed GNR pairs grown at different *h*-BN stack depth (see Extended Data Fig. 9b) forming a vertical GNR/*h*-BN/GNR junction at the crossing point that can be harnessed to fabricate tunnelling devices with an atomically-precise tunnelling barrier[50]; and (iii) continuous planar Y-shaped GNR junctions (see Extended Data Fig. 9c) that may serve as a basis for three-terminal devices, such as side-gate field effect transistors.[13] The new catalytic growth approach presented herein constitutes a crucial step towards gaining control over the *in-situ* growth of such advanced encapsulated architectures, which is expected to reshape the field of GNR based nano-electronics. Moreover, the presence of spin-polarized topological edge states along zigzag edges holds promise for spintronic and quantum computing devices.

*Methods*

**GNR growth.** *h*-BN flakes were mechanically exfoliated onto commercial $SiO_2$/Si substrate using the scotch tape method. The exfoliated flakes were exposed to hydrogen plasma at 300 °C to remove all residual organic matter. Catalytic iron nanoparticles were deposited on the *h*-BN covered $SiO_2$/Si surfaces through thermal evaporation (evaporation rate: ~0.01 nm/s, base vacuum pressure: ~$1\times10^{-6}$ mbar). The decorated chips were then put into a 1-inch quartz tube furnace (Anhui BEQ Equipment Technology) and flushed with a mixture of hydrogen and argon for air evacuation. After that, the system was gradually heated up to the growth temperature (800-900 °C) under hydrogen and argon gas mixture ($H_2$:Ar = 1:1) at atmospheric pressure. Upon reaching growth temperature, the argon atmosphere was replaced by methane to commence GNR growth. After a growth period of 5-60 min, the systems were cooled down to room temperature under a protective hydrogen and argon atmosphere ($H_2$:Ar = 1:1). As shown in Fig. 1a and b, at a growth temperature of ~850°C, with methane as a carbon source, embedded GNR growth can be initiated at the nanoparticle surface into one of the *h*-BN interlayer gaps. The SEM images taken on an *h*-BN flake before (Extended Data Fig. 2a) and after (Extended Data Fig. 2b) CVD growth demonstrate this growth process. Before growth, the *h*-BN substrate is featureless, whereas after growth, GNRs extending from the *h*-BN edges and steps appear (straight bright lines).

The growth of GNRs has not been observed on *h*-BN flakes thinner than 5 nm. This lack of growth on very thin *h*-BN flakes can be attributed to the fact that they inherit the roughness of the underlying SiO$_2$/Si substrate (see SI Fig. S8), thus eliminating the superlubric sliding of the growing GNRs, which is crucial for their elongation. For thicker *h*-BN stacks, this roughness is screened away from the substrate by the elastic response of the layers.

Comparing to GNR growth on *h*-BN surfaces which exhibits a wide temperature window of 600-900°C, the embedded GNR growth typically requires higher temperatures and exhibits a relatively narrow temperature window of 800-900 °C. Consequently, at temperatures below 800 °C, we only observe on-surface GNR growth. Conversely, at temperatures ranging from 800 to 900 °C, we observe GNR growth both on *h*-BN surfaces and between *h*-BN layers.

**Hydrogen-plasma etching.** As stated above, the CVD growth process yields not only embedded GNRs, but also on-surface grown GNRs and CNTs. To remove the latter on surface structures, following the growth process, the samples were put into a quartz tube of a plasma-enhanced chemical vapor deposition (PECVD, Anhui BEQ Equipment Technology) system, where they were heated to an etching temperature of 300 °C under hydrogen for 30 min. Then the samples were further etched by hydrogen plasma (RF power 30 W, ~45 Pa) for 15 min. The embedded GNRs were protected from the moderate hydrogen plasma exposure by the inert *h*-BN stack. Extended Data Fig. 3 demonstrates the removal of two GNRs from the *h*-BN surface by plasma treatment, with no effect on the embedded GNRs.

**Scanning electron microscopy.** The high throughput large-scale mapping capability of low-energy scanning electron microscopy (SEM), which is commonly employed to characterize CNTs, can be readily harnessed for the rapid mapping and characterization of our embedded-GNRs. After plasma etching, large scale SEM imaging was performed at the *Inlens SE* mode of a Zeiss SEM system (model: GeminiSEM 300). A low electron energy (0.5-1 keV) was used to obtain good GNR contrast.

**Scanning transmission electron microscopy.** The cross-section TEM specimens were prepared by *in-situ* lift-out via milling in a ESCAN FE-SEM (GAIA3 GMU Model 2016) equipped with a focused ion beam (FIB) apparatus. Before milling, thick (~ 2 μm) protective platinum was deposited over the section of interest, such that the *h*-BN stacks were sandwiched between the Pt coating and the $SiO_2$ substrate. This allowed us to safely cut the sample and image the embedded GNRs. The cross-section specimens were fabricated with a starting milling condition of 30 kV/500 pA and then stepped down to 2 kV/20 pA to minimize the ion-beam induced damage. High-resolution imaging of specimen cross-sections was performed in a FEI Talos F200X STEM at 200 kV using high-angle annular dark-field (HAADF) and dark field (FEI DF4) detectors. The HAADF detector had a collection angle of 6-34 mrad for Z (atomic number) contrast imaging. The DF4 detector had a collection angle of 2-5 mrad for integrated differential phase contrast (iDPC) imaging. A beam current of 1 pA, and a camera length of 1.1 m were used for STEM image acquisition.

**Atomic force microscopy.** AFM measurements were performed using a Cypher S AFM system (Asylum Research). AFM probes of AC200 and RTESPA-300 were typically used for the imaging in AC topography mode in air. For high resolution AFM scanning in AC mode, PFQNE-AL and RTESPA-150 probes were used. For lattice resolution *h*-BN imaging the friction mode was used, and the lattice structure and orientation were obtained via fast Fourier transform.

**Device fabrication and electrical characterization.** To make electrical contacts, both ends of an embedded GNR were exposed by reactive ion etching (RIE), with the etching windows defined by standard e-beam lithography (GeminiSEM 300, Zeiss) and PMMA acting as an etching mask. The sample was etched in a RIE system using plasma generated from a mixture of $CHF_3$ and $O_2$ with a flow rate of 40 SCCM and 6 SCCM, respectively, with RF power of 60 W under a pressure of 10 Pa (or using $SF_6$ plasma, flow rate: 60 SCCM, power: 20 W, pressure: 13 Pa). Then, metal electrodes (5 nm Cr/50 nm Au) are deposited using an e-beam evaporator. Electrical transport measurements were conducted using two Keithley 2400 source-meters.

**Density functional theory (DFT) calculations.** All the DFT calculations were performed using the Vienna *ab initio* simulation package (VASP)[51-53] with the projector augmented wave (PAW) method.[54] The PBE generalized gradient approximation for the exchange-correlation energy density functional was adopted.[55] Van der Waals interactions were accounted for using the DFT-D3 approach.[56] The plane-wave cutoff energy was set to 650 eV, and the Brillouin zone was sampled using a Monkhorst-Pack k-mesh with a separation criterion of 0.02.[57] Criteria for energy and force convergence were set to $10^{-4}$ eV and $10^{-2}$ eV/Å, respectively.

**Force field structural optimization protocol.** Armchair and zigzag GNR segments of ~3 nm width and ~5 nm length were placed in the middle of multilayer *h*-BN stacks of 6, 20, 30, and 40 layers (SI Fig. S1). The edges of the GNRs were passivated by hydrogen atoms to avoid peripheral C-C bond reconstruction, that may influence friction.[58] The intra-layer interactions within the GNRs and the *h*-BN layers were computed using the second generation REBO potential[59] and the Tersoff potential,[60] respectively. The interlayer interactions between the GNRs and the *h*-BN layers were described via the registry-dependent interlayer potential (ILP)[61-63] with refined parametrization,[58,64] as implemented in LAMMPS.[65] The models were optimized using the FIRE algorithm[66] with a threshold force value of $10^{-6}$ eV Å$^{-1}$. These force-fields were also used to perform the moiré pattern, potential energy surface, and GNR trajectory calculations and for the frictional molecular dynamics simulations described below.

**Sliding potential energy surface calculations.** To calculate GNR sliding potential energy surfaces (PESs) appearing in Fig. 3c-e, armchair and zigzag GNR segments (~3 nm width and ~5 nm length) were placed on a monolayer *h*-BN substrate or within an AA′-stacked *h*-BN bilayer. The isolated GNR segments were first relaxed using the optimization protocol described above. The optimized GNRs were then placed atop an *h*-BN layer (of experimental lattice parameter[62] of 2.505 Å) or between two such layers, with a graphene/*h*-BN distance of 3.3 Å. The lateral center-of-mass position of each GNR was then further adjusted by performing rigid-body zero temperature NVE ensemble simulations for 0.15 ns with fixed *h*-BN surface(s). PESs were then obtained by rigidly shifting the optimized GNR segments in the lateral directions and recording the energy variations based on the classical potentials described above. The PESs are normalized by the number of carbon atoms in the GNRs.

**Sliding trajectory simulations**. To obtain the sliding trajectories appearing in Fig. 3c-e, we performed further structural optimization of the GNRs used for the sliding energy landscape calculations (see above), allowing for lateral atomic position relaxation using the FIRE algorithm with a threshold force value of $10^{-6}$ eV Å$^{-1}$. The GNRs were then forced to slide by applying a constant lateral force on the carbon atoms at the first row of the trailing edge of the $3 \times 5$ nm$^2$ GNR, keeping the *h*-BN layers fixed. The forces used were of 2.0 nN for the embedded AC-GNR applied in the armchair direction of the *h*-BN layers, and 0.3 or 7.6 nN for the embedded or on-surface ZZ-GNR applied in the zigzag direction of the *h*-BN layers. Excess heating was avoided by applying velocity damping to all GNR atoms at a rate of 1 ps$^{-1}$. To conform with the lateral sliding PESs appearing in Fig. 3c-e, only lateral motion of the GNR atoms was allowed during these simulations. The center-of-mass position was recorded yielding the trajectories presented in (Fig. 3c-e). Very similar sliding trajectories were obtained when allowing also for vertical motion of the GNR atoms between the fixed *h*-BN surfaces (see SI Fig. S9).

**Penetration depth simulations.** The detailed protocol for the penetration depth simulations appearing in Fig. 3a, b is described in SI Section 15.

**Estimation of mobility and subthreshold swing of GNR devices.** The carrier mobility $\mu$ of the embedded-GNR FETs was estimated based on standard transistor models via $\mu = \frac{dG}{dV_g} \times L/C_{gs}$, where $G$ is the conductance of the FET device, $L$ is the channel length, and $C_{gs}$ = 10.05 pF/m is the effective capacitance. Here, $\frac{dG}{dV_g}$ was obtained by linear fitting of the transfer characteristic curves, as shown in Extended Data Figs. 7a and 7b. The effective capacitance $C_{gs}$ was calculated through three-dimensional electrostatic simulations, performed using the Fast Field Solvers software (http://www.fastfieldsolvers.com/). The simulated structure included a large back-gate plane; a dielectric layer (representing the *h*-BN crystal and the underlying SiO$_2$ substrate) with the same lateral dimension as the back-gate, 285 nm thickness, and a dielectric constant of $\varepsilon_0$=3.9; a 3 nm wide GNR lying above the dielectric layer; and two metal fingers representing the contacts with the corresponding experimental dimensions. In order to get precise results, a cartesian grid spacing of ~1 nm was used. Classical considerations suggest that the geometry of the object and the dielectric constant of the medium determine the capacitance. Additional quantum capacitance contributions can be safely neglected in our case.[67] Once $C_{gs}$ was calculated, we used the above relation to evaluate the room temperature carrier mobilities yielding values as high as 4,600 cm$^2$V$^{-1}$s$^{-1}$ (Extended Data Figs. 7a and 7c) and 2,155 cm$^2$V$^{-1}$s$^{-1}$ (Extended Data Figs. 7b and 7d), respectively. We note that the carrier mobility extracted via the relation $\mu = \frac{1}{\rho n e}$, where $\rho$ is the resistivity, $n$ is the charge carrier density, and $e$ is the electron charge, is consistent with the values extracted using the field-effect method (see SI section 18 for details). While the carrier mobility analysis of our embedded GNR FET device presented above and the high on-off ratio ($I_{on}/I_{off}$ ~ 10$^6$) exhibited by the as-measured transfer curves appearing Fig. 4c are important parameters for the device characterization they are by far not the only figures of merit relevant for logic transistors. Another important factor is the form of the transfer characteristics near and below threshold, which is governed by the electrostatic nature of the transistor. This can be quantified by the subthreshold swing parameter stating by how many mV the gate voltage has to be varied to change the drain current by one decade. The embedded-GNR FET devices that we fabricated exhibit subthreshold swing values as low as ~100 mV/dec (obtained by linear fittings of the semi-log transfer curves, as shown in Extended Data Figs. 7f and 7g), which is close to the theoretical room temperature lower bound of 60 mV/dec for conventional FETs.[68]

**Acknowledgments**

This work is supported by the National Key R&D Program of China (no. 2021YFA1202902, 2020YFA0309000, 2022YFA1405400, 2022YFA1402702), the National Natural Science Foundation of China (no. 12374292, 12074244, 12102307, 11890673, 11890674, 11874258, 12074247, 12174249 and 92265102), the open research fund of Songshan Lake Materials Laboratory (no. 2021SLABFK07). W.O. acknowledges the Natural Science Foundation of Hubei Province (2021CFB138) and the start-up fund of Wuhan University. M.U. acknowledges the financial support of the Israel Science Foundation (grant no. 1141/18) and the ISF-NSFC joint grant 3191/19. O.H. is grateful for the generous financial support of the Israel Science Foundation (grant no. 1586/17), Tel Aviv University Center for Nanoscience and Nanotechnology, the Naomi Foundation via the 2017 Kadar Award, and the Heineman Chair of Physical Chemistry. Shiyong Wang acknowledges support from Shanghai Municipal Science and Technology Qi Ming Xing Project (no. 20QA1405100), Fok Ying Tung Foundation for young researchers. K.W. and T.T. acknowledge support from the Elemental Strategy Initiative conducted by the MEXT, Japan, (grant no. JPMXP0112101001, JSPS KAKENHI (grant nos. 19H05790 and 20H00354) and A3 Foresight by JSPS. Shiyong Wang and Zhiwen Shi acknowledges support from SJTU (21X010200846), and additional support from the Shanghai talent program. L.Q. acknowledges the support from the Basic Research Laboratory Support Program (grant no. 2021R1A4A1033224) of the National Research Foundation of Korea. B.L. acknowledges support from the Development Scholarship for Outstanding Ph.D. of Shanghai Jiao Tong University. We would also like to acknowledge support from the Instrument Analysis Center of Shanghai Jiao Tong University (SJTU) for performing FIB on GAIA3, and STEM on TALOS F200X. Molecular dynamics simulations were carried out at the National Supercomputer TianHe-1(A) Center in Tianjin and the Supercomputing Center of Wuhan University.

**Author contributions**

B.L. and Z.S. initiated the project. Z.S., M.U., F.D., O.H., and W.O. supervised the project. B.L., J.C. and S.L. grew the samples. J.C. and B.L. carried out TEM and STEM measurements. B.L., J.C. and P.S. carried out SEM measurements. B.L., J.C. and S.L. carried out AFM measurements. P.S., B.L. and J.C. fabricated devices and conducted electron transport measurements. Sen Wang and J.X. conducted MD simulations of GNR sliding. W.O. designed the MD simulation setup and implemented the codes. L.Q., I.M. and F.D. carried out theoretical calculations. K.W. and T.T. grew *h*-BN single crystals. B.L., J.C., S.L., Sen Wang, P.S., L.Q., I.M., C.L., C.H., X.Z., W.O., J.X., X.W., J.J., Q.L., Shiyong Wang, G.C., T.L., M.U., O.H., F.D., and Z.S. analysed the data. B.L., J.C., Sen Wang, S.L., L.Q., I.M., W.O., M.U., O.H., F.D., and Z.S. wrote the paper with input from all authors.

**Data availability**

The data supporting the findings of this study are available within this paper and its Supplementary Information files or from the corresponding authors upon request.

**Code availability**

The code related to the findings of this study are available from the corresponding authors upon request.

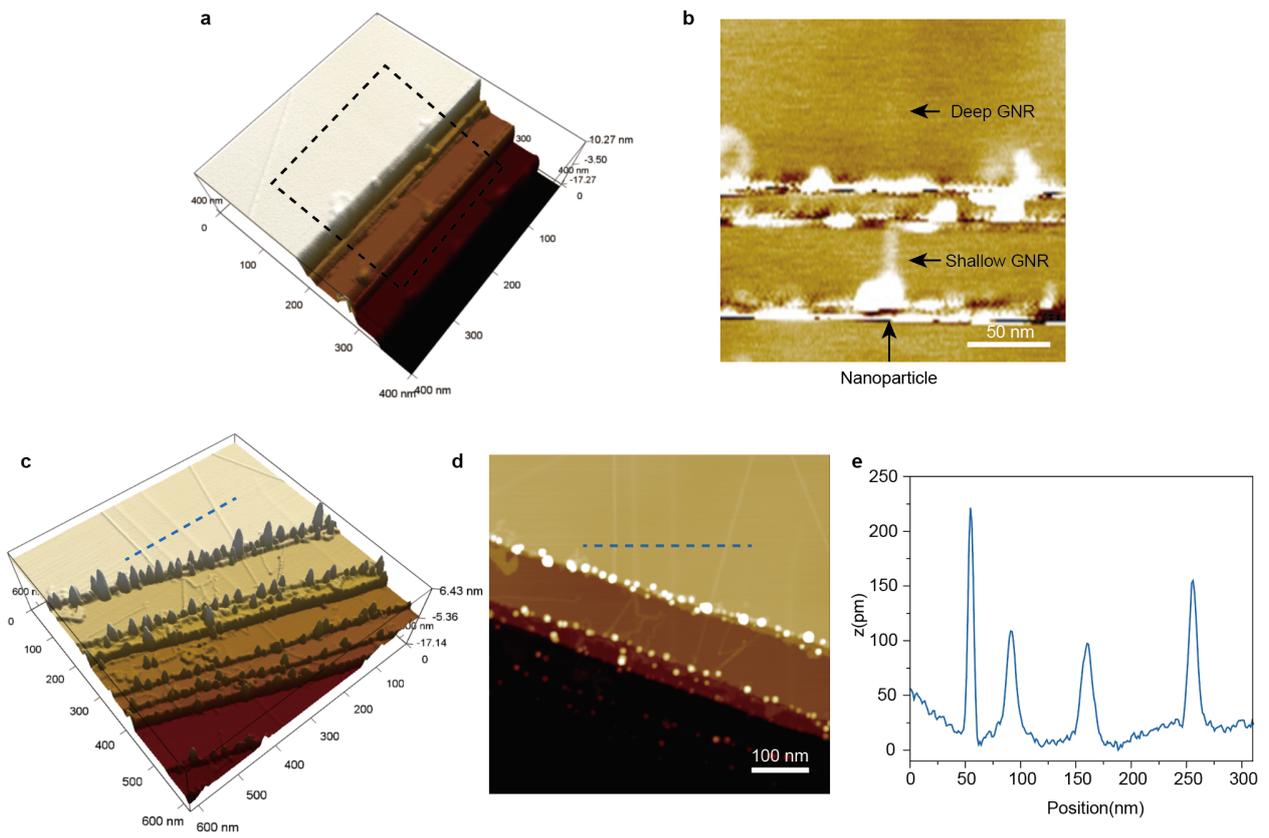

**Extended Data Figure 1 | AFM topography images of embedded GNRs and Fe nanoparticles at *h*-BN step edges. a,** 3D AFM topography image of *h*-BN step edges after Fe nanoparticle deposition and CVD growth. **b,** Zoom-in on the region marked by the dashed square in panel (**a**) demonstrating an embedded-GNR grown from a nanoparticle into the *h*-BN stack. **c** and **d,** AFM topography images of *h*-BN step edges with high nanoparticle density. **e,** Height profile taken along the blue dashed line in panel (**c**) and (**d**).

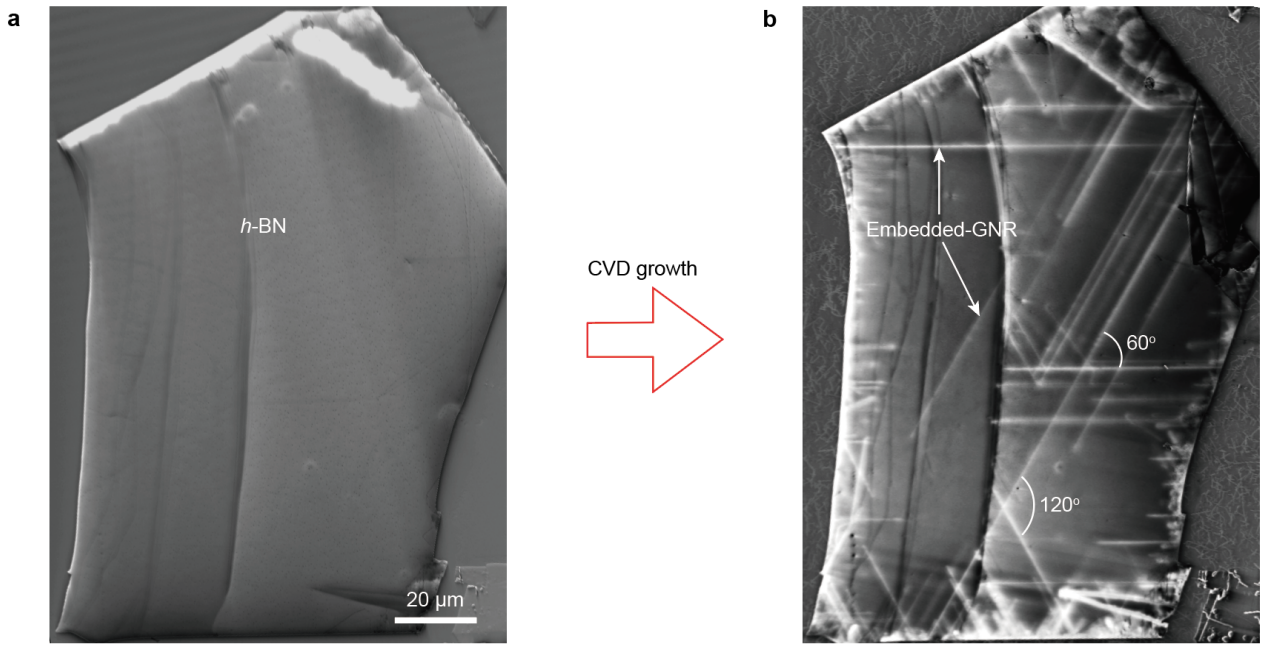

**Extended Data Figure 2 | Demonstration of CVD growth of embedded GNRs. a,** SEM image of a bare *h*-BN sample prior to growth. **b,** SEM image of the same sample as in panel a, following CVD growth. The bright lines are SEM fingerprints of embedded GNRs grown from edge-positioned catalytic nanoparticles.

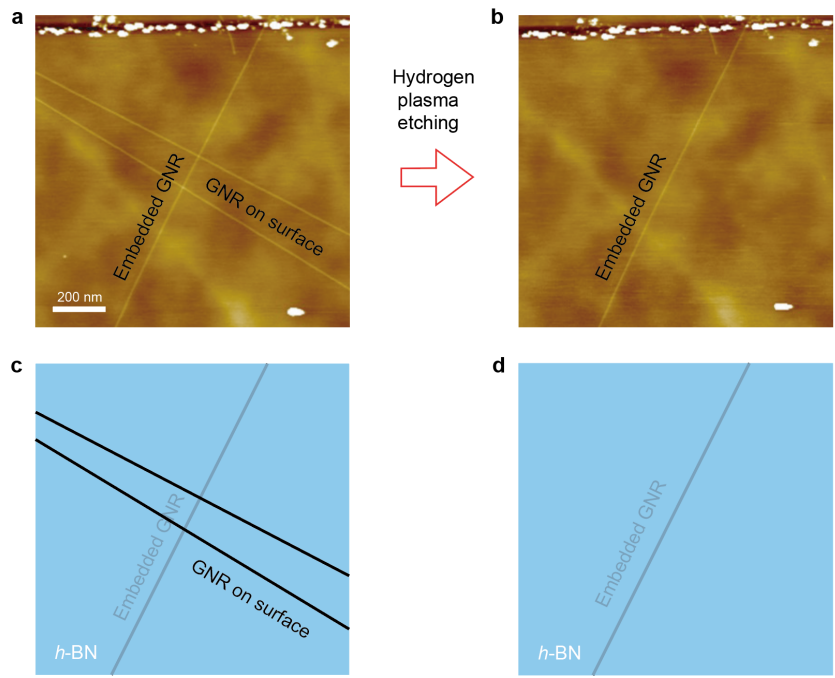

**Extended Data Figure 3 | Removal of on-surface GNRs through plasma etching.** Upper two panels are AFM topography images captured at the same *h*-BN surface region following CVD growth, before **a** and after **b** plasma etching, respectively. Panels **c** and **d** present a schematic view of panels **a** and **b**, respectively.

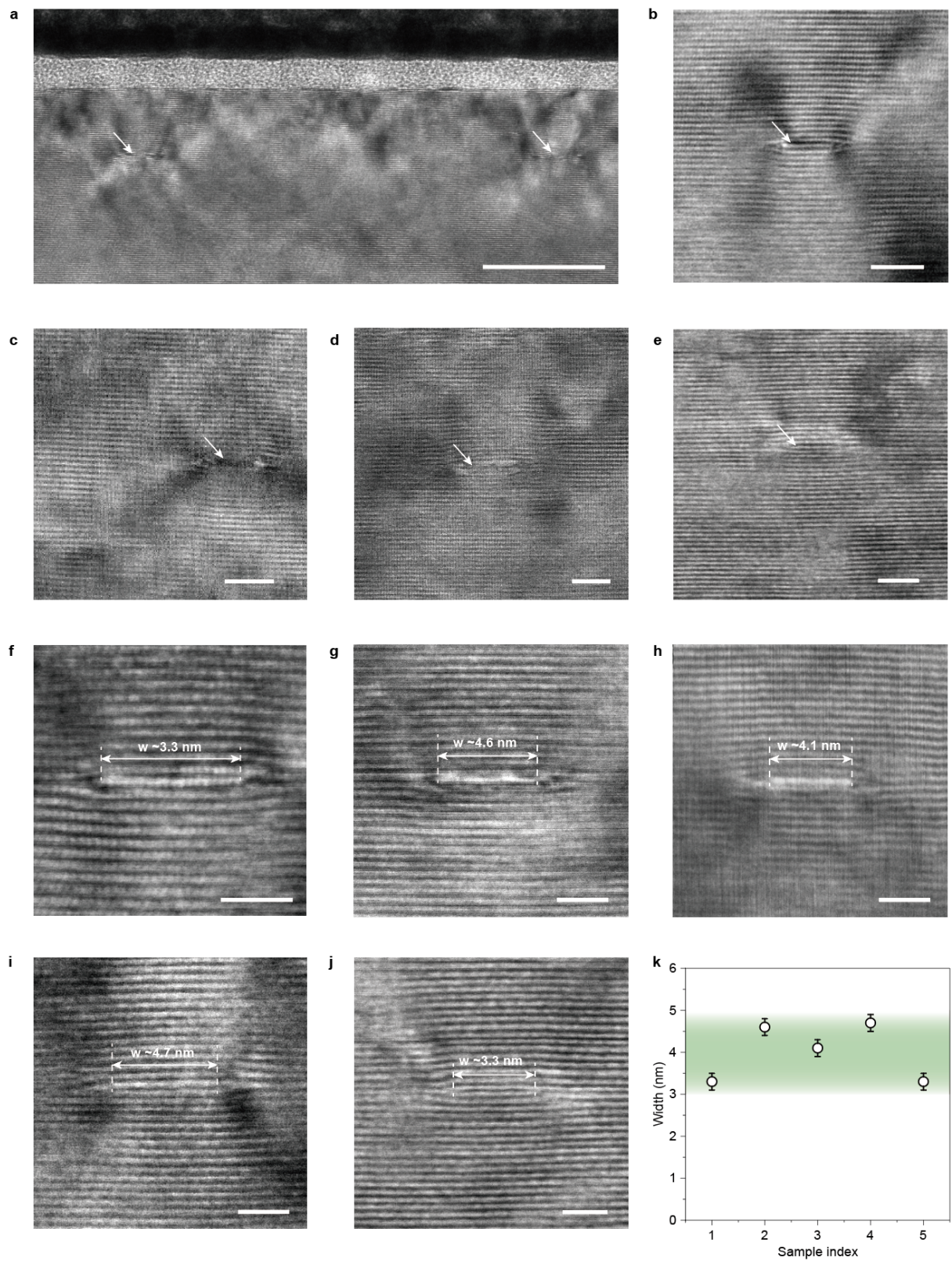

**Extended Data Figure 4 | Additional cross-sectional STEM images of embedded GNRs. a,** A large-scale STEM side-view image showing two embedded GNRs. **b-e,** Zoom-in STEM images. **f-j,** High-resolution STEM images of five embedded-GNRs with width in the range from 3 nm to 5 nm. **k,** Width statistics. Scale bar: (a) 20 nm, (b-e) 3 nm, (f-j) 2 nm.

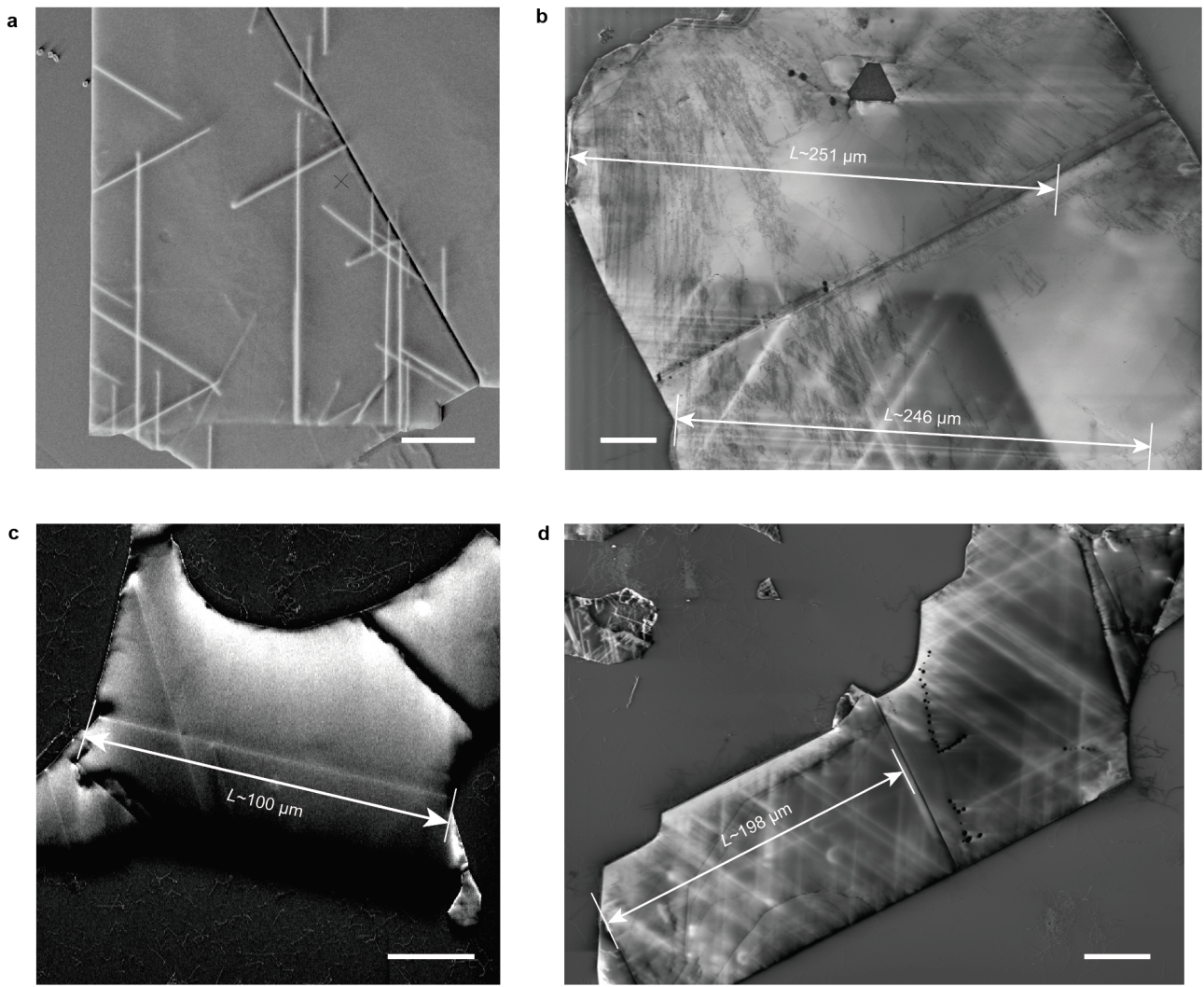

**Extended Data Figure 5 | Large-scale SEM images of embedded GNRs.** Straight embedded-GNRs of typical length of a few tens of micrometers mostly oriented along three distinct directions on the *h*-BN flake that are angularly separated by 60°. Scale bar: (a) 5 μm, (b) 30 μm, (c) 20 μm and (d) 40 μm.

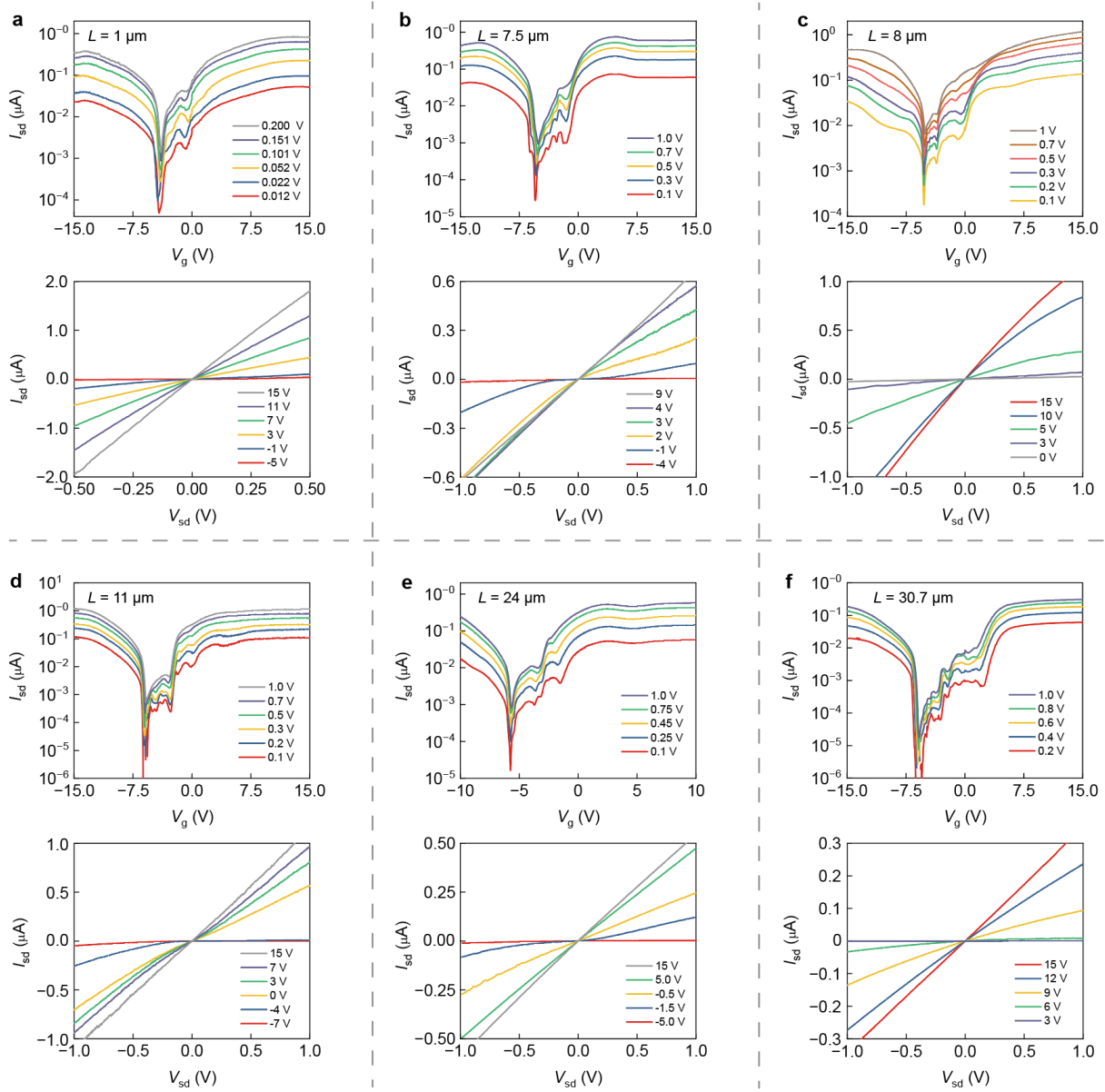

**Extended Data Figure 6 | Transfer and output characteristics of additional GNR devices. a-f,** Room temperature transfer and output characteristics of six devices with channel lengths ranging from 1 μm to 30.7 μm. See SI section 7 for a discussion of the oscillations observed in the transfer curves.

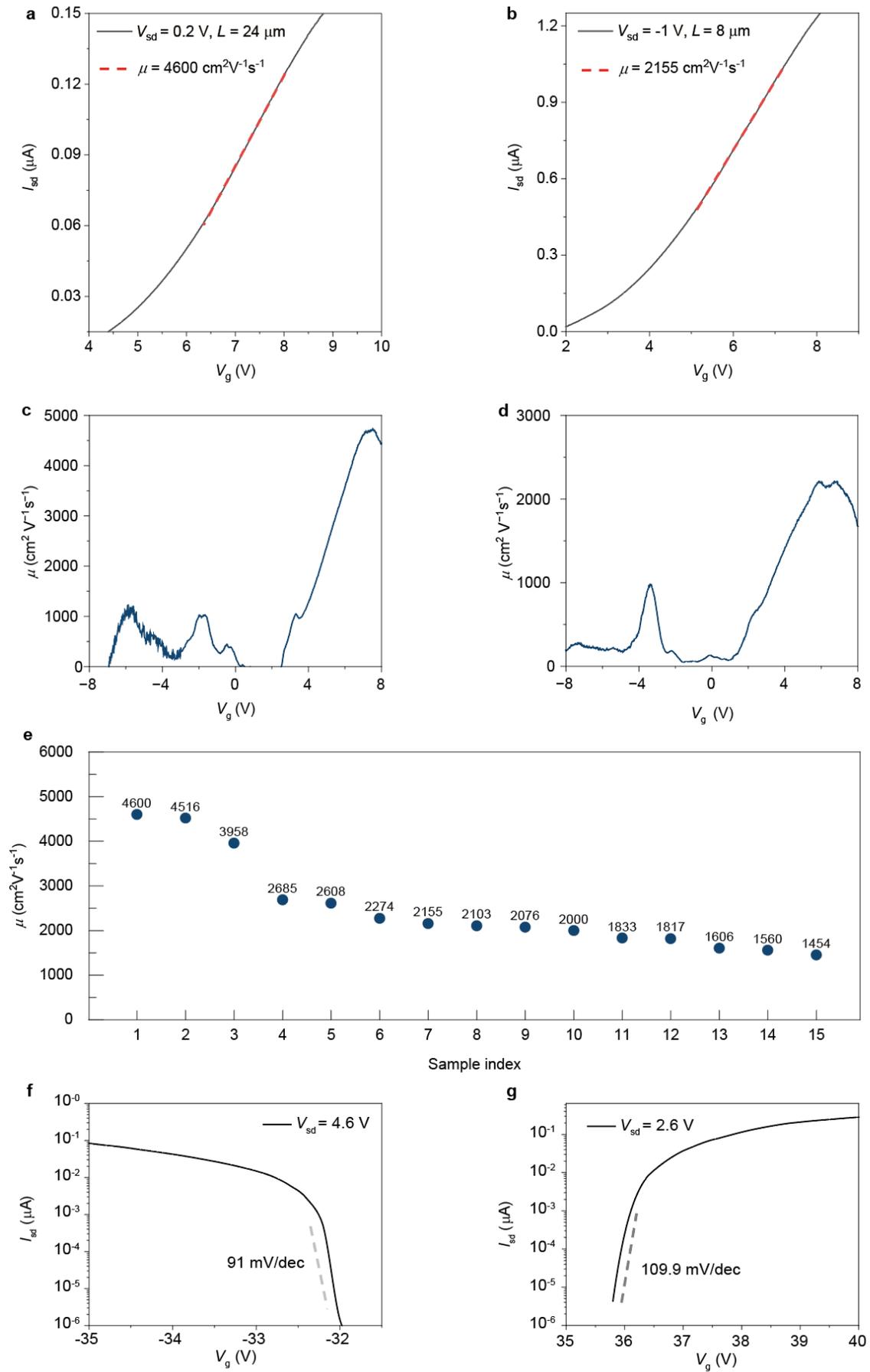

**Extended Data Figure 7 | Estimation of mobility and subthreshold swing of GNR devices.**

**a,** Room temperature transfer characteristics (black) of a GNR device of channel length of $L$

~24 μm, width of ~3 nm, and SiO$_2$ thickness of 285 nm ($C_{gs}$= 10 pF/m). A linear fit (red dashed line) yields a carrier mobility of ~4,600 cm$^2$V$^{-1}$s$^{-1}$. **b,** Room temperature transfer characteristics (black) of a different device ($L$ ~8 μm, width of ~3 nm, and SiO$_2$ thickness of 285 nm, $C_{gs}$ = 10 pF/m), demonstrating a carrier mobility of ~2,155 cm$^2$V$^{-1}$s$^{-1}$. **c** and **d,** Carrier mobility as a function of gate voltage extracted from the transfer characteristics of the devices shown in panel (**a**) and (**b**), respectively. **e,** The statistics of carrier mobility. Semi-log plots of the transfer characteristic curves of a (**f**) 17 μm and (**g**) 8 μm long channel devices (~3 nm in width) demonstrating similar room temperature subthreshold swing values of ~ 100 mV/dec, as extracted from the slope of the dashed grey lines.

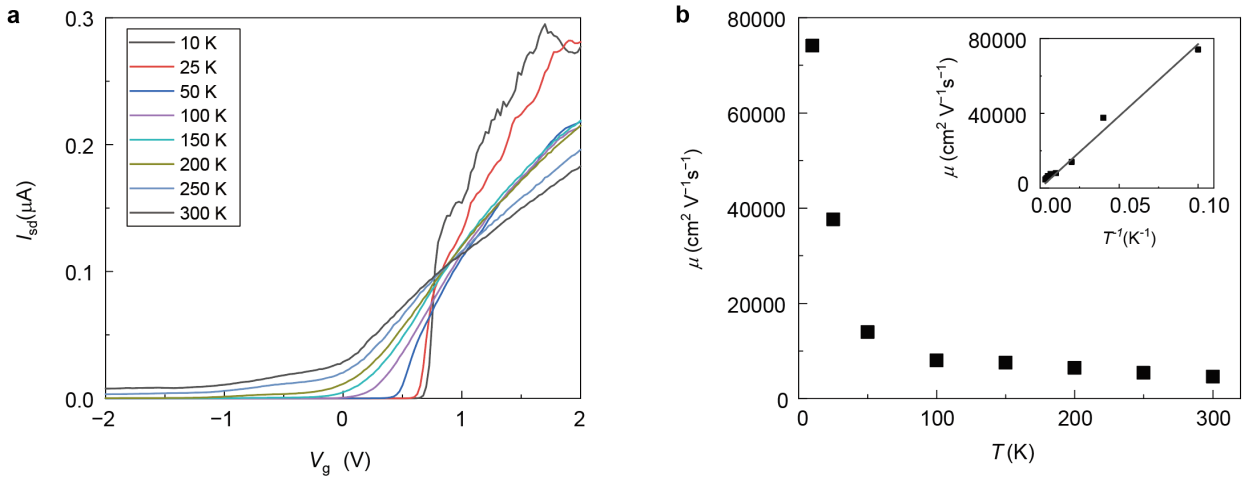

**Extended Data Figure 8 | Temperature-dependent carrier mobility. a,** Transfer characteristics of a high-performance device at $V_{sd}$ = 0.5 V for different temperatures. **b,** Measured mobility as a function of temperature. Inset: the measured mobility as a function of $T^{-1}$ (black squares) and the corresponding linear fit.

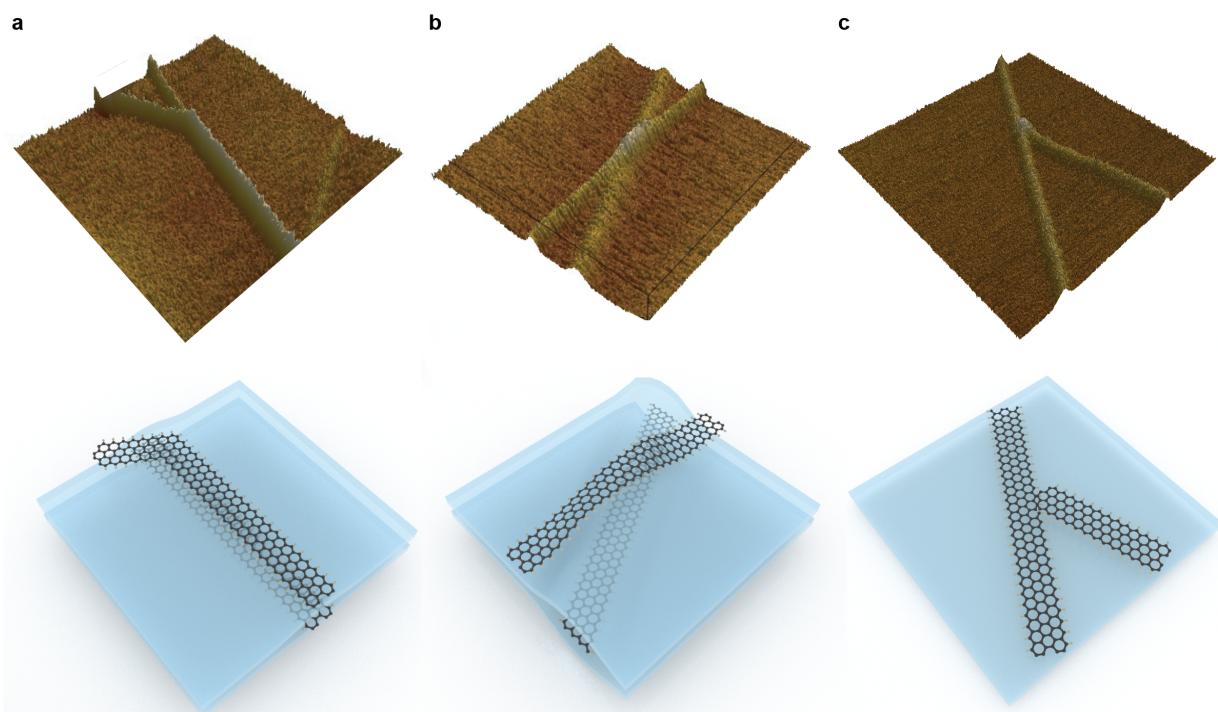

**Extended Data Figure 9 | Representative types of inter-ribbon architectures.** Three-dimensional AFM height images (upper panels) and schematic illustrations (lower panels) of (**a**) vertically parallel, (**b**) vertically crossed, and (**c**) intersected planar Y-junction GNR architectures. For clarity, the top *h*-BN layers are hidden in the three schematic illustrations.